\documentclass[a4paper]{jpconf}
\usepackage{graphicx}
\usepackage{color}
\usepackage[usenames,dvipsnames]{xcolor}
\usepackage{amsmath}
\usepackage{amssymb}
\usepackage{cite}
\graphicspath{{figures/}}

\newcommand{\pdagger}{{\phantom{\dagger}}}
\newcommand{\neel}{N\'{e}el}
\newcommand{\dt}{\Delta\tau}
\newcommand{\dT}{\Delta\tau}
\newcommand{\iwn}{i \omega_n}
\newcommand{\reff}[1]{Fig.\ \ref{fig:#1}}

\newcommand{\reffb}[1]{Fig.\ \ref{fig:#1} (b)}

\newcommand{\refq}[1]{(\ref{eq:#1})}

\renewcommand{\vec}[1]{\mathbf{#1}}
\newcommand{\vk}{\vec{k}}
\newcommand{\myparagraph}[1]{{\it #1} -- }

\begin{document}
\title{Deciding the fate of the false Mott transition in two dimensions by exact quantum Monte Carlo methods}

\author{D. Rost$^{1,2}$ and N. Bl\"{u}mer$^1$}
\address{$^1$ Institute of Physics, Johannes Gutenberg University, Mainz, Germany}
\address{$^2$ Grad. School Materials Science in Mainz, Johannes Gutenberg University, Mainz, Germany}
\ead{rostda@uni-mainz.de}

\begin{abstract}
We present an algorithm for the computation of unbiased Green functions and self-energies for quantum lattice models, free from systematic errors and valid in the thermodynamic limit. The method combines direct lattice simulations using the Blankenbecler-Scalapino-Sugar quantum Monte Carlo (BSS-QMC) approach with controlled multigrid extrapolation techniques. We show that the half-filled Hubbard model is insulating at low temperatures even in the weak-coupling regime; the previously claimed Mott transition at intermediate coupling does not exist.
\end{abstract}

\section{Introduction}
Numerous studies of the Hubbard model \cite{Hubbard_1959} have greatly enhanced our understanding of strongly correlated electron systems within the last four decades. 
Specialized methods, namely the semi-analytic Bethe ansatz and the density matrix renormalization group (DMRG) \cite{DMRG}, yield reliable high-precision results (only) in the case of one spatial dimension.
Conversely, the dynamical mean-field theory (DMFT) \cite{Metzner, Georges_1996} provides deep insight in the limit of high dimensionality.
However, the intermediate regime of two (and three) dimensions is less well understood, e.g., with respect to pseudogap physics \cite{Lee2006, Armitage2010, Rost2012} and high-$T_\text{c}$ superconductivity.
In this paper, we will address another open question: the nature of the Mott metal-insulator transition (MIT) \cite{MIT1, MIT2} of the half-filled Hubbard model in two dimensions ($D=2$). 

As indicated by shaded regions in \reff{PhaseDiagram}, the Hubbard model is insulating at half filling (1 electron per site) and sufficiently strong coupling, in any dimension, while it is metallic at weak coupling -- as long as the translational symmetry is not broken, e.g., by magnetic order.
For $D=3$, one finds antiferromagnetic (AF) ordering \cite{Staudt2000}, which implies insulating behavior, below the \neel \ temperature $T_N$ (gray dash-dotted line).
Such long-range order is ruled out by the Mermin-Wagner theorem in $D=2$ (at temperature $T>0$).
Single-site DMFT predicts a MIT with critical interactions $2.3 \lesssim U_c \lesssim 2.9$ (pink dotted line).
This transition line is shifted to weaker interactions within cluster DMFT (CDMFT), which includes short-range correlations on small clusters (violet solid line) \cite{Park2008}. 
However, the low-$T$ metallic phase implied by the CDMFT result is incompatible with several earlier studies highlighting effects of ``short-range antiferromagnetism'' \cite{Gorelik2012,Rost2012,Chang2013}.
To clarify the situation, we apply the Blankenbecler-Scalapino-Sugar quantum Monte Carlo (BSS-QMC) \cite{BSS_1981} approach with controlled multigrid extrapolation techniques to selected points in the phase diagram (green crosses). 
We will show that these points are separated by a MIT line (or narrow crossover) and conclude that the $2D$ Mott physics is quite similar to the $3D$ case \cite{Fuchs2011}, in spite of the Mermin-Wagner theorem.

\section{Model and Methods}

The single-band Hubbard Hamiltonian is given by
\begin{equation}\label{eq:Hubbard}
  \hat{H} =\! - \ t \ \sum_{\langle ij\rangle ,\sigma}  \hat{c}^{\dag}_{i\sigma}
\hat{c}^\pdagger_{j\sigma} 
+   U\sum_{i}  \, \left(\hat{n}_{i\uparrow}-\frac{1}{2}\right)\left( \hat{n}_{i\downarrow}-\frac{1}{2} \right) \text{,}
\end{equation}
with number operators $\hat{n}_{i\sigma}=\hat{c}^{\dag}_{i\sigma}\hat{c}^\pdagger_{i\sigma}$, next-neighbor hopping $t$, and local interactions $U$. 
In this paper, we set $t=0.25$ and consider finite-size clusters with periodic boundary conditions.
 
\begin{figure}[t]
	\centering
	\begin{minipage}[b]{0.47\columnwidth}
		\centering
		\includegraphics[width=1.0\columnwidth]{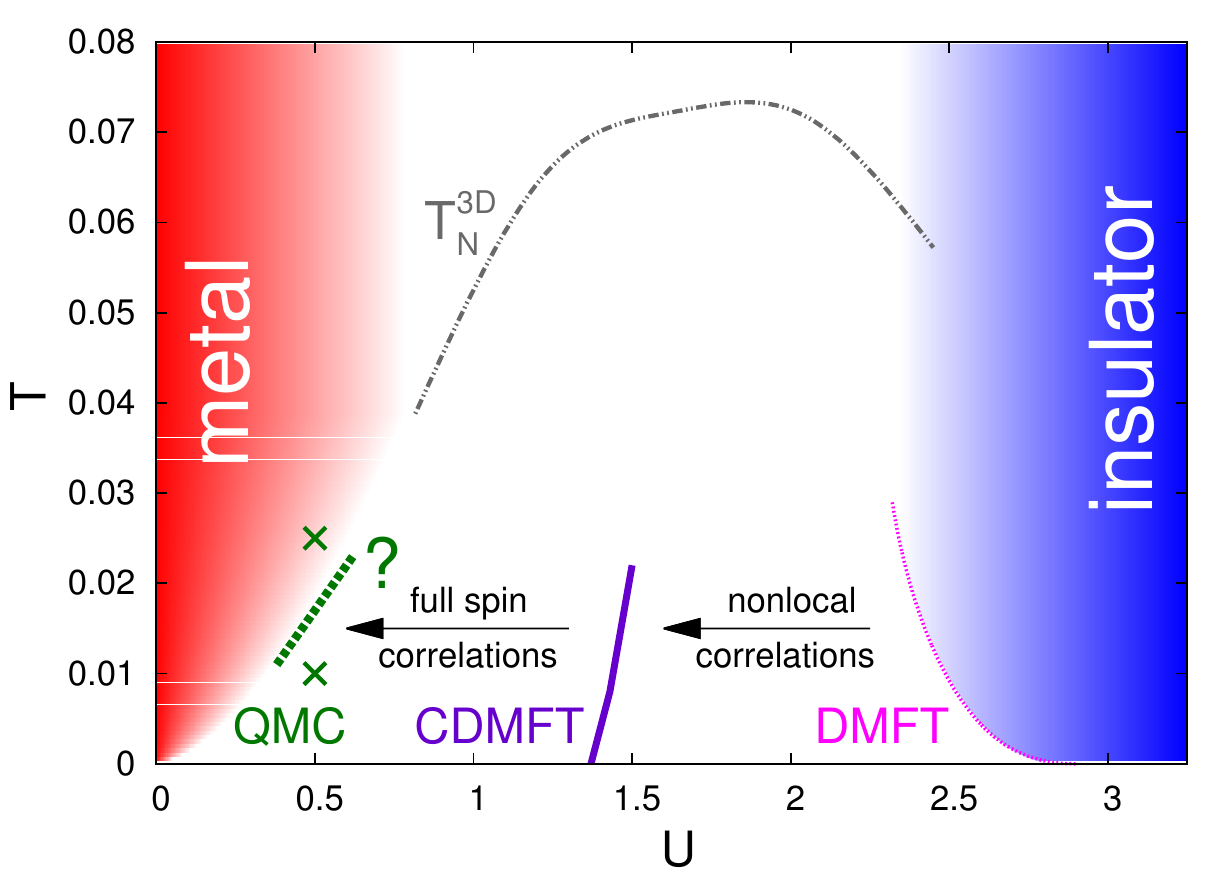}
		\caption{Schematic representation of the MIT in the Hubbard model at half filling on a square lattice. Green crosses: parameters selected for our unbiased QMC study. Gray dash-dotted line: \neel \ temperature for the cubic lattice (rescaled).}
		\label{fig:PhaseDiagram}
	\end{minipage}
	\begin{minipage}{0.03\columnwidth}
	\phantom{??}
	\end{minipage}
	\begin{minipage}[b]{0.47\columnwidth}
		\centering
		\includegraphics[width=1.0\columnwidth]{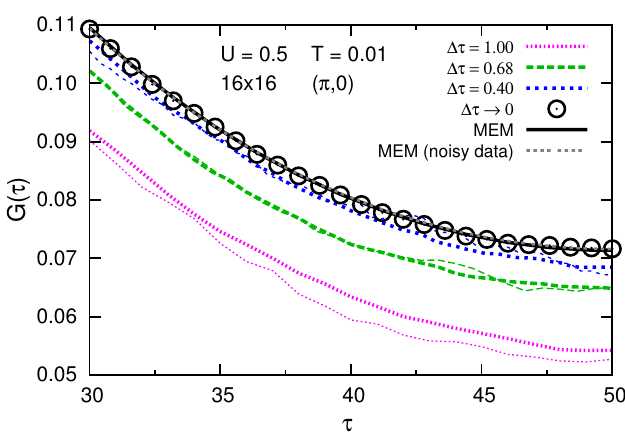}
		\caption{Extrapolation of BSS-QMC Green functions to $\dt \rightarrow 0$.
		 Analytic continuation (MEM, black line) regularizes the results of pointwise extrapolations (symbols) and yields stable results (gray line) even in the case of noisy raw data (thin color lines).}
		\label{fig:Gtau_smooth_16x16}
	\end{minipage}
\end{figure}

The BSS-QMC algorithm is based on a Trotter-Suzuki decomposition of the partition function \begin{equation}
  \label{eq:trotter_decomp}
     Z = \, \text{Tr} \left( e^{-\beta (H_t + H_U)} \right) \approx \, Z_{\dt} =  \text{Tr} \left( \prod_{l=0}^\Lambda e^{-\dT H_t}e^{-\dT H_U} \right)  \text{,}
\end{equation}
where $H_U$ ($H_t$) corresponds to the interaction (kinetic) term in \refq{Hubbard}, $\beta = 1/T$ is the inverse temperature ($k_b \equiv 1$) and $\Lambda$ denotes the number of time slices ($\Lambda = \beta / \dt$). 
A discrete Hubbard-Stratonovich transformation simplifies the interaction term, which is quartic in the fermionic operators, to a quadratic form and a coupling to an auxiliary  Ising field $h$. As usual for quadratic Hamiltonians, the fermionic degrees of freedom can be integrated out and the partition function is expressed by determinants of matrices:
\begin{align}\label{eq:trotter_decomp3}
  Z_{\dt} &= \sum\nolimits_{\{ h \}} \text{det} \left[ M_\uparrow^{\{ h \}} \right] \text{det}\left[ M_\downarrow^{\{ h \}} \right]  \text{.}
\end{align}
The sum in \refq{trotter_decomp3}, and finally the computation of all observables, is performed by Monte Carlo methods.  
The main results of the BSS-QMC algorithm are Green functions $G_{ij}(\tau_l)$ on a discrete imaginary-time grid $\{\tau_l \in [0,\beta]\}$ for each pair of real-space coordinates $\{i,j\}$ on the given lattice. The involved matrix operations for the calculation of the determinants in \refq{trotter_decomp3} lead to a scaling of $\mathcal{O}(\Lambda \, N_c^3)$, 
which restricts the algorithm to relatively small clusters ($N_c\lesssim 400$).

To arrive at reliable conclusions regarding the Mott transition using BSS-QMC, one has first to eliminate all systematic errors (finite-size as well as Trotter errors) and, secondly, calculate the self-energy on the Matsubara axis, $\Sigma (\iwn)$, from the imaginary-time Green function in a stable manner. These steps are described in the following.

\section{Elimination of systematic errors}
As the elimination of Trotter errors is well established \cite{Bluemer2007, Bluemer2008, Gorelik2009, Rost2013}, we will sketch the scheme only briefly and focus on the problem specific and physically more relevant finite-size analysis and present a stable method to calculate unbiased Green functions, even if the raw data are noisy.


\myparagraph{Trotter bias} 
The raw BSS-QMC Green functions (thick colored lines in \reff{Gtau_smooth_16x16}) do not only live on different imaginary-time grids $\{ \tau_i \}$, depending on the chosen discretization $\dt$, but are also shifted with respect to each other (and to the exact solution). After aligning them on a common fine grid \cite{fn:splines} the Trotter errors can be eliminated with high precision by piecewise extrapolations of $\dt\to 0$ (circles) \cite{Rost2013}. However, some fluctuations remain inevitably, especially when using noisy raw data (using a factor of 10 less QMC sweeps; thin colored lines). These can be greatly reduced by regularization via a maximum entropy method (MEM), which computes corresponding spectral functions via the inversion of 
\begin{equation}
  \label{eq:Aw}
	G(\tau) = - \int_{-\infty}^\infty d\omega \, A(\omega) \, \frac{e^{-\omega \tau}}{1 + e^{-\omega \beta}} \, \text{ .}
\end{equation}
Note that the intermediate spectra $A(\omega)$ are used only for producing continuous and smooth Green functions $G(\tau)$ via \refq{Aw}; in this combination, the procedure is stable.
After this step, the results based on high- or low-precision data (thick black line vs.\ gray line) are hardly distinguishable and can both be used for stable computations of self-energies.


\begin{figure}[t]
	\includegraphics[width=0.5\columnwidth]{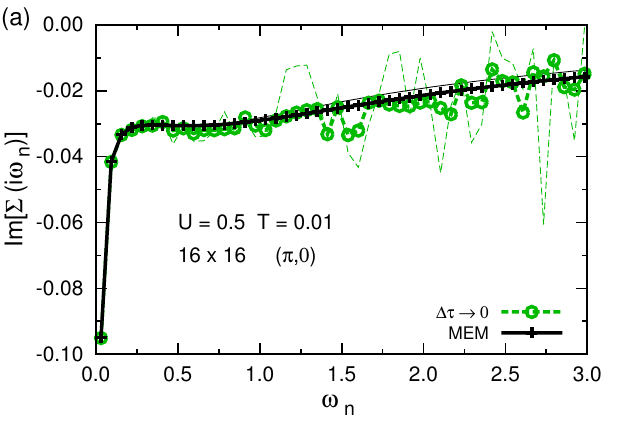}
	\includegraphics[width=0.5\columnwidth]{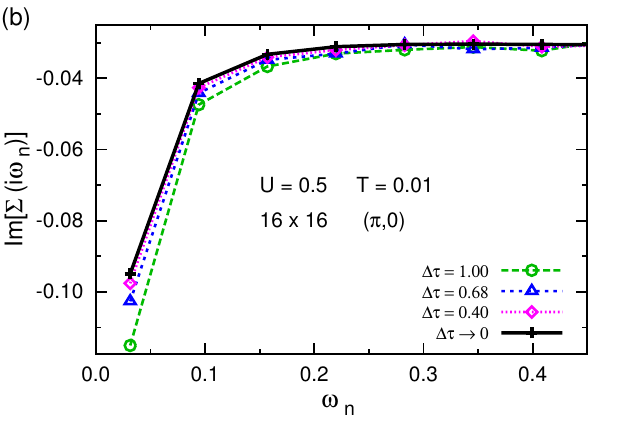}
	\caption{(a) Influence of statistical noise in the raw data $G(\tau)$ on the self-energy $\Sigma(i\omega_n)$ and stabilization by MEM regularization. Thin lines: noisy data, thick lines and symbols: high precision results. (b) Impact of Trotter discretization $\dt$ on the Matsubara self-energy.}
\label{fig:SE_smooth_16x16}
\end{figure}

\myparagraph{Self-energy}
These quasi-continuous $G(\tau)$ can be reliably Fourier transformed to the Matsubara axis:
\begin{equation}
  \label{eq:tau_iwn}
	G(\iwn) = \int_0^\beta d\tau \, G(\tau)\, e^{-\iwn \tau} \text{ .}
\end{equation}
A quantity of great interest for the analysis of the Mott transition is the imaginary part of the momentum-resolved self-energy, which is connected to the BSS-QMC Green function  $G$ and the non-interacting Green function $\mathcal{G}$ via a Dyson equation \cite{Fetter_Walecka_1971}
\begin{equation}
  \label{eq:dyson}
	\Sigma^{\phantom{-}}_{\vec{k}}(\iwn) =  \mathcal {G}_{\vec{k}}^{-1}(\iwn) - G_{\vec{k}}^{-1}(\iwn) = \iwn + \mu - \epsilon^{\phantom{-}}_{\vec{k}} - G_{\vec{k}}^{-1} \, \text{ ,}
\end{equation}
where $\mu$ is the chemical potential and $\epsilon_\vk$ the dispersion of the noninteracting problem. 
As shown in \reffb{SE_smooth_16x16}, the MEM regularizing procedure leads to estimates (crosses and black line) of the imaginary part of the self-energy (the real part vanishes at $\vk=(\pi,0)$ due to particle-hole symmetry) which are smooth even at higher frequencies. In contrast, the direct results (green circles) fluctuate significantly even when based on high-precision raw data (and even more so for noisy raw data; thin dashed green lines). Note that both methods agree very well at the lowest Matsubara frequencies, where the QMC predictions are most reliable.

In \reffb{SE_smooth_16x16} the influence of the Trotter discretization is shown. The Trotter biased self-energies (broken colored lines) deviate from the exact result (black solid line) for large values of $\dt$, in particular at the lowest frequencies. However, in the parameter regime explored in this study, we can afford small enough values of $\dt$ so that the elimination of Trotter errors is less essential than the finite-size extrapolation to be discussed below.


\begin{figure}[t]
	\includegraphics[width=\columnwidth]{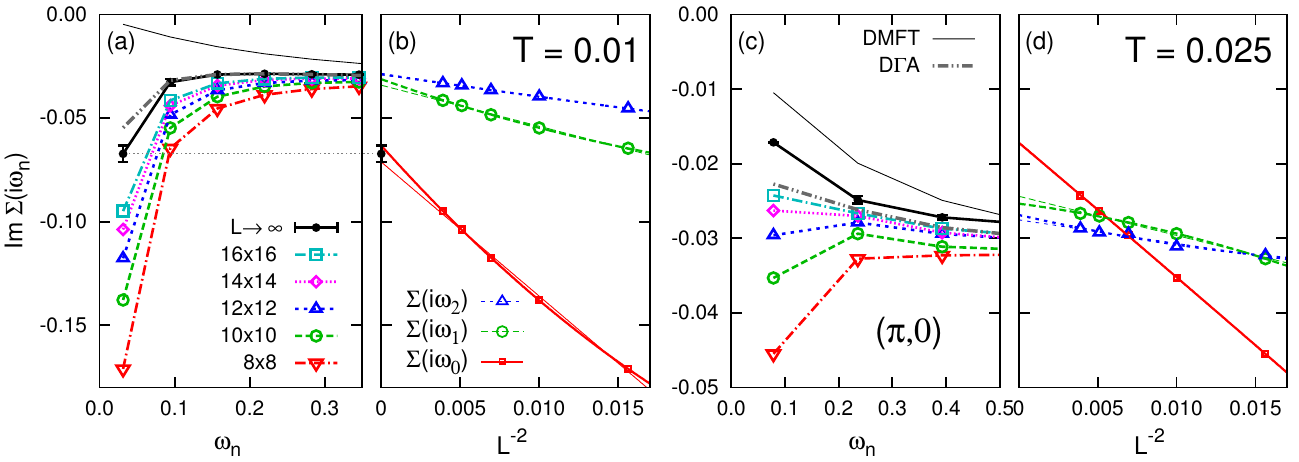}
	\caption{Finite-size scaling of Matsubara self-energy 
	at $U = 0.5$, $\mathbf{k}=(\pi,0)$.  Finite-size BSS-QMC data (open symbols and broken colored lines), extrapolated BSS-QMC results in the thermodynamic limit (circles and black bold solid line), and D$\Gamma$A data (gray line) versus Matsubara frequency $\omega_n$ at $T=0.01$  (a) and $T=0.025$ (c); also shown are momentum-independent single-site DMFT results (thin black lines).  (b)+(d) Finite-size BSS-QMC (symbols) data for the first three Matsubara frequencies versus inverse system size plus extrapolations in linear order in $L^{-2}$ (thin lines) and quadratic order (thick lines).}
\label{fig:SE_L}
\end{figure}

\myparagraph{Finite-size extrapolation of the self-energy}
Using the methods described above, unbiased and stable estimates of the Matsubara self-energy have been obtained for square $L\times L$ clusters with sizes ranging from $8\times 8$ to $16 \times 16$ (with a computational effort that varies as $N_c^3=L^6$, i.e., by a factor of $2^6=64$ in this range). Results at the ``anti-nodal'' \cite{fn:nodal} momentum point $\vk=(\pi,0)$ [throughout the paper, unit lattice spacing $a=1$ is assumed] are shown (colored symbols and broken lines) in \reff{SE_L}(a) for the low temperature $T=0.01$ and in \reff{SE_L}(c) for the elevated temperature $T=0.025$, respectively. Evidently, the finite-size effects are enormous: while the smallest systems ($8\times 8$, triangles) show insulating behavior, i.e., a strong enhancement of $\Sigma(i\omega_n)$ towards small frequencies, at both temperatures, this tendency is reduced with increasing $L$ at $T=0.01$ [\reff{SE_L}(a)] and completely eliminated at $T=0.025$ [\reff{SE_L}(c)]. Obviously, careful extrapolations are needed for reliable predictions in the thermodynamic limit.

These are, indeed, possible, as shown in \reff{SE_L}(b) for $T=0.01$ and in \reff{SE_L}(d) for $T=0.025$, respectively, for the three lowest Matsubara frequencies (where the finite-size effects are largest): as a function of $L^{-2}$, 
the finite-size results (symbols) can be fitted with second-order polynomials (thick lines) with high precision; as these fit functions have small curvatures, linear extrapolations (thin lines) to the thermodynamic limit (i.e., $L^{-2}\to 0$) deviate only slightly from the quadratic ones. We use these deviations as error bars and the arithmetic average of both extrapolation procedures as final result, as indicated for $\Sigma(i\omega_0)$ by a black symbol in \reff{SE_L}(b).

\section{Mott physics and correlation lengths}
These final QMC results [black symbols and lines in \reff{SE_L}(a) and \reff{SE_L}(c)] show that the character of the system changes drastically between the selected phase points (crosses in \reff{PhaseDiagram}): while the QMC self-energy indicates a metallic phase at $T=0.025$, very similar to the DMFT solution [thin black line in \reff{SE_L}(c)], the low-temperature phase is clearly insulating - and completely unlike the paramagnetic DMFT solution  [thin black line in \reff{SE_L}(a)]. In contrast, the QMC results are strikingly similar to the D$\Gamma$A predictions (gray dash-dotted lines), especially at low $T$.
The dynamical vertex approximation ($D\Gamma A$) \cite{DGA1, DGA2}  is a diagrammatic extension of DMFT which works directly in the thermodynamic limit (without any need for finite-size extrapolations); for the full set of $D\Gamma A$ results see \cite{Schaefer2014}. QMC and D$\Gamma$A together show convincingly that the suspected phase transition (dotted line in \reff{PhaseDiagram}) is, indeed, present.

The question remains how exactly this metal-insulator transition (or crossover) is related to magnetic properties. After all, long-range AF order is excluded, in $D=2$, by the Mermin-Wagner theorem. However, as shown in \reff{corr}(a), the spin correlations decay much more slowly at $T=0.01$ (circles) than at $T=0.025$ (squares). More precisely, the correlation length at the elevated temperature can be estimated (using $16\times 16$ QMC data) to $\xi\approx 3.7$ while it is clearly too large ($\xi \gg 30$) at $T=0.01$ to allow precise determination using QMC. For this observable, D$\Gamma$A predictions, shown in \reff{corr}(b) and \reff{corr}(c) are more reliable, yielding correlation lengths of $\xi\approx 4$ at the higher temperature, in excellent agreement with QMC, and $\xi >300$ at $T=0.01$, also consistent with QMC. Using D$\Gamma$A we could also show that the temperature dependence of $\xi$ changes its character at the Mott transition (see \cite{Schaefer2014}).

\begin{figure}[t]
	\includegraphics[width=0.65\columnwidth]{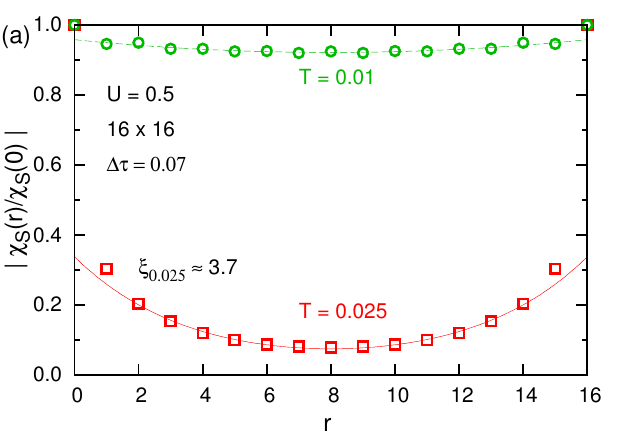}
	\includegraphics[width=0.35\columnwidth]{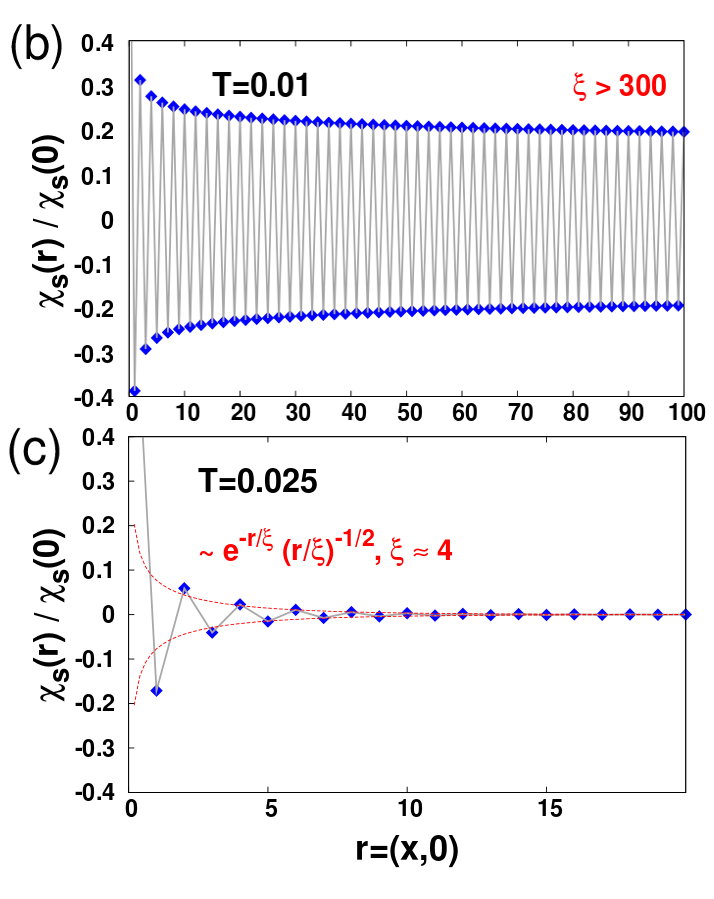}
	\caption{Real-space spin correlation function. (a) The correlation length for $T=0.025$ (red squares) from BSS-QMC calculations for a $16 \times 16$ lattice is estimated by an exponential fit (red solid line). For $T=0.01$ (green circles) $\xi$ is much larger than the extend of the lattice. \textit{Right panel:} The corresponding estimations from D$\Gamma$A data \cite{Schaefer2014} for $T=0.01$ (b) and $T=0.025$ (c) confirm the BSS-QMC results. }
\label{fig:corr}
\end{figure}

\section*{Conclusion and acknowledgements}
In spite of its simplifications, the Hubbard model remains a challenging target for theorists, especially in the interesting case of two spatial dimensions. All available methods have certain limitations, many of which are of particular concern in the nonperturbative intermediate-coupling regime at low temperatures and in the presence of longer-range correlations. Therefore, a single method can hardly yield authoritative results; some predictions may even be completely off (such as the CDMFT transition line in \reff{PhaseDiagram}).

In the study presented in this work (with focus on the QMC methodology; for the full story, see \cite{Schaefer 2014}), we have applied two methods (an unbiased variant of BSS-QMC as well as the dynamical vertex approximation) with completely different characteristics. Only the near-perfect agreement between both sets of results makes the predictions truly authoritative (and validates technical choices on either side). 
We have settled one important question at half filling, namely the character of the Mott metal-insulator transition on the square lattice: it is driven by exponentially long-ranged \cite{Schaefer 2014} antiferromagnetic correlations, which act similarly to the AF long-range order in the cubic case, and is {\bf not} connected to a quantum-critical point at $U>0$.

The same methodology may be useful in other parameter ranges, e.g., for frustrated or anisotropic Hubbard models, doped systems, or multi-band models. It might also be worthwhile to compute transport properties, within bubble approximation or beyond, in order to observe the metal-insulator transition even more directly (than via the self-energy). 
  
\vspace{2ex}
We acknowledge support from the research unit FOR 1346 of the German Research Foundation (DFG) and the graduate school GSC 266.


\section*{References}

\newcommand{\PRA}{\textit{Phys. Rev.} A }
\newcommand{\PRB}{\textit{Phys. Rev.} B }
\newcommand{\PRD}{\textit{Phys. Rev.} D }
\newcommand{\PRE}{\textit{Phys. Rev.} E }
\newcommand{\EPJB}{\textit{Eur. Phys. J. B }}
\newcommand{\PTPS}{\textit{Prog. Theor. Phys. Suppl.\ }}

\end{document}